\documentclass[10pt]{article}
\usepackage[utf8]{inputenc}
\usepackage[T1]{fontenc}
\usepackage{amsmath,amssymb,bm}
\usepackage[version=4]{mhchem}
\usepackage{graphicx}
\usepackage{authblk}
\usepackage{placeins}
\usepackage[numbers,sort&compress]{natbib}
\usepackage[hidelinks]{hyperref}
\usepackage{xcolor}

\newcommand{\added}[1]{#1}

\title{Origin and Reduction of Coercive Fields in ZnO-based Wurtzite Ferroelectrics}

\author[1,3]{Abhijeet Dhakane}
\author[2]{Alireza Sepehrinezhad}
\author[5]{William Prudnick}
\author[5]{Jon-Paul Maria}
\author[1,3]{Bobby Sumpter}
\author[4]{Adri C. T. van Duin}
\author[3]{Kyle P. Kelley}
\author[1,3]{P. Ganesh}

\affil[1]{Bredesen Center for Interdisciplinary Research, University of Tennessee, Knoxville, TN 37996, USA}
\affil[2]{Department of Mechanical Engineering, The Pennsylvania State University, University Park, PA 16802, USA}
\affil[3]{Center for Nanophase Materials Sciences, Oak Ridge National Laboratory, Oak Ridge, TN 37831, USA}
\affil[4]{Department of Mechanical Engineering, The Pennsylvania State University, University Park, PA 16802, USA}
\affil[5]{Department of Materials Science and Engineering, The Pennsylvania State University, University Park, PA 16802, USA}
\affil[*]{Correspondence and requests for materials should be addressed to A.D. \added{(adhakane@alum.utk.edu)}, K.P.K. (kelleykp@ornl.gov) or P.G. (ganeshp@ornl.gov).}
\date{}

\begin{document}
\maketitle

\begin{abstract}
{Wurtzite ferroelectrics offer a route to semiconductor-compatible non-volatile devices, but their large coercive fields (E$_c$) constrain the operating window between polarization reversal, leakage and dielectric breakdown. Atomically localized and non-classical switching fronts have recently been identified in wurtzite nitrides, yet the microscopic feature that sets the coercive field remains unresolved. Here we use large-scale reactive molecular dynamics, parameterized against first-principles data, to follow field-driven reversal in pristine, Mg-modified and heterostructured ZnO. Reversal proceeds through rugged inversion-boundary filaments in agreement with recent observations in other wurtzite ferroelectrics. We find that the field required for switching is controlled by the advancing filament head, which is a under-coordinated reconstructed inversion-boundary region during the transient switching dynamics with a large field-aligned local polar-order amplitude that is only partially compensated by an oppositely oriented surrounding shell. Controlled strain-only and charge-doping-only perturbations show that charge redistribution lowers the coercive field more effectively than strain because it suppresses this field-setting unscreened local polar order at the filament head, with the combined effect lowering E$_c$ by $\sim$ 50 \%.  Layered ZnO/ZnMgO architectures further create buried nucleation sites and shorten filament propagation lengths. These results connect atomistic switching topology to a materials-design principle for substantially reducing coercive fields in ZnO-based wurtzite ferroelectrics.}
\end{abstract}

\section*{Introduction}

\added{Ferroelectric thin films are being re-positioned as active materials for embedded non-volatile memories, ferroelectric field-effect electronics, neuromorphic circuits, electro-optic and nonlinear photonic platforms, microelectromechanical systems, and high-temperature electronics\cite{zu2023large, pradhan2024scalable, li2025ferroelectric, mikolajick2023ferroelectric} . This shift has been enabled by two materials families that are more compatible with semiconductor processing than conventional perovskite oxides. Fluorite \ce{HfO2}-based ferroelectrics showed that switchable polarization can persist in nanoscale CMOS-compatible films \cite{boscke2011ferroelectricity,park2015ferroelectricity,lee2020scale,pesic2016fieldcycling}. Wurtzite ferroelectrics then extended ferroelectricity to polar semiconductors and nitrides, beginning with \ce{AlScN} and expanding to \ce{AlBN}, \ce{ZnMgO}, and heterostructured wurtzites \cite{fichtner2019alscn,liu2021aluminum,liu2021post,pradhan2024scalable,casamento2024perspectives,fichtner2025polarization}. Their wide band gaps, large spontaneous polarizations, high thermal stability, optical transparency, piezoelectric response, and compatibility with sputtering or atomic-layer deposition make them attractive as integrated ferroelectric semiconductors for electronic, electromechanical, and photonic technologies. Their limitation is because coercive fields remain too large for many practical low-voltage devices, often approaching the margin set by leakage and dielectric breakdown. From a device perspective, the central challenge is therefore not only whether a wurtzite film can switch, but whether its polarization can be written, read, retained, cycled, and spatially patterned under practical operating fields \cite{scott1989ferroelectric,mikolajick2021next,chen2016review,wang2025review}. The key unresolved issue is what local structural--electrostatic feature sets the coercive field and how that feature can be engineered to lower switching fields in wurtzite ferroelectrics for integrated electronic, electromechanical, and photonic devices.}

Mg-substituted ZnO is an ideal platform for answering this question because it begins from a simple binary polar semiconductor whose polarization is normally difficult to reverse. Experiments have established robust ferroelectric hysteresis, wake-up and composition-dependent switching in \(\mathrm{Zn}_{1-x}\mathrm{Mg}_x\mathrm{O}\) thin films \cite{ferri2021ferroelectrics,jacques2023wake,spurling2025composition,aronson2025aldzmo}. First-principles studies have linked switchability to competition between the polar wurtzite lattice, Mg-induced local distortions, metastable nonpolar coordination environments and the proximity of competing phases \cite{huang2022origin,zu2023large,malashevich2007first,moriwake2014wurtzite,moriwake2020lowcoercive}. ZnMgO should therefore not be viewed as a simple average-composition alloy. Its response depends on the distribution of local environments created by composition, strain, disorder, electronic redistribution, interfaces and processing. ZnO/ZnMgO based heterostructures further show that buried interfaces and proximity effects can activate switching in otherwise hard-to-switch polar layers \cite{skidmore2025proximity,eliseev2025proximity,baksa2024strain,sepehrinezhad2026heterostructures}. These observations point to a local origin of coercive-field reduction, but they do not yet identify the microscopic descriptor that should be optimized.

{The broader wurtzite-ferroelectric literature has established that polarization reversal is intrinsically non-classical. In a wurtzite lattice, switching requires the metal sublattice to cross the basal anion plane and reconstruct its tetrahedral coordination, rather than simply displacing within an approximately fixed coordination cage \cite{konishi2016mechanism,lee2024switching}. Atomic-resolution studies of AlScN and AlBN have identified inversion boundaries and characteristic 4--8 topological motifs in switched material \cite{wolff2021atomic,calderon2023atomic,hayden2021albn,hayden2023albnsi,calderon2024albnlocal}, while kinetic measurements show thermally activated, nucleation-controlled reversal \cite{zhu2021temperature,yazawa2023abrupt,lu2024domain}. The recent Rappe-group study of AlN further demonstrated that reversal can advance through rapid one-dimensional cation-column cascades and produce a fractal-like lateral domain front \cite{behrendt2026fractals}. Complementary in situ STEM measurements on epitaxial (Al,B,Sc)N reveal a different mesoscale morphology, one where domains nucleate near an electrode, develop zigzag and inclined boundaries.  At atomic resolution, the boundary itself exhibits a structurally complex configuration consistent with either projection through an inclined wall or an extended non-polar $\beta$-BeO-like region \cite{calderon2026domain}. Together, these observations indicate that the detailed switching morphology depends on material chemistry, disorder, interfaces and electrostatic boundary conditions rather than following a single universal geometry.}

{This diversity sharpens the central materials question. Identifying a filamentary or domain-wall pathway explains \emph{how} reversal can propagate, but does not by itself identify \emph{what sets the field required to initiate and advance that pathway}. A predictive low-field switching design rule requires a field-setting descriptor, i.e.  a local structural--electrostatic feature whose energetic cost co-varies with the coercive field and can be deliberately modified by chemistry or architecture.}

This distinction is essential because several mechanisms can produce similar switching signatures. Global strain can lower ideal barriers, epitaxial mismatch can reduce the field in AlScN, and thickness scaling can reduce operating voltage in conventional ferroelectrics \cite{zakutayev2021reduced,jiang2022ultralow}. Defects, dead layers and interfacial domain walls can either assist switching by nucleating domains or impede it through pinning and charge trapping \cite{hwang2024idb,wang2025deadlayer,pesic2016fieldcycling}. Dopants may act through size mismatch, local strain, charge redistribution, chemical bonding, screening or heterogeneous nucleation. Without a descriptor that separates these effects, coercive fields are usually optimized empirically by changing composition, growth conditions or electrode stacks. A predictive strategy must identify which part of the moving switching nucleus carries the dominant electrostatic penalty, and then use chemistry and architecture to lower that penalty.

{Resolving this problem requires simulations large enough to connect local nucleation to collective domain evolution. Density-functional-theory and nudged-elastic-band calculations identify local barriers and metastable coordination environments, but accessible cells are generally too small to resolve finite-temperature filament connectivity, lateral coalescence and interactions among multiple switching fronts. Continuum and phase-field approaches reach larger length scales but necessarily coarse-grain the atomic reconstruction at inversion boundaries. Large-scale reactive molecular dynamics occupies the intermediate regime needed here: it retains atomistic bonding, alloy disorder and topological reconstruction while permitting switching fronts to evolve over tens of nanometres \cite{dhakane2023graph,vanduijn2001reaxff,senftle2016reaxff,akbarian2019reaxff,sepehrinezhad2024reaxff,thompson2022lammps}.}

{Here we compare pristine ZnO, uniformly Mg-modified ZnO, controlled charge-only and strain-only perturbations, and layered ZnO/ZnMgO/ZnO architectures using a Zn/Mg/O ReaxFF model developed for switchable ZnO-based ferroelectrics. ZnO reverses through inversion-boundary filaments, but the principal result is not the existence of the filamentary pathway itself. The advancing filament has a reconstructed core--shell structure. during its transient dynamics.  The head of the filament core is under-coordinated and develops a large field-aligned local polar-order amplitude while an oppositely oriented surrounding shell partially compensates it. Across controlled perturbations, the coercive-field reduction follows suppression of this field-setting local order more directly than structural strain alone. The same mechanism suggests a second design lever---buried ZnO/ZnMgO interfaces that provide internal nucleation sites and reduce the distance over which the high-cost filament head must propagate. We therefore use the switching trajectory to derive a local design strategy, which is to reduce the electrostatic cost of the advancing inversion-boundary head and engineer where such heads nucleate, leading to a {\em dynamics-by-design} strategy for discovering Wurtzite materials that would switch at lower coercive fields.}

\section*{Results}

\subsection*{Pristine ZnO separates the switching pathway from the field-setting event}

\begin{figure}[!p]
  \centering
  \includegraphics[width=\linewidth]{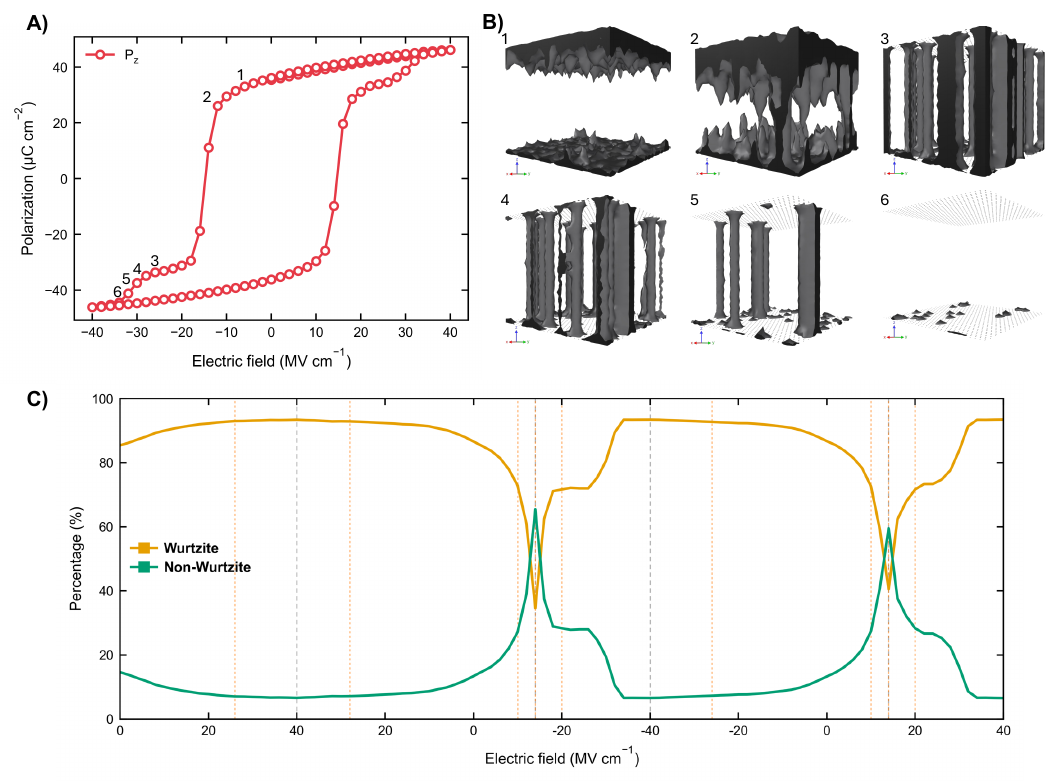}
  \caption{{\textbf{Field-driven reversal in pristine ZnO proceeds through inversion-boundary filaments.} \textbf{A}, Simulated polarization--field hysteresis loop with a coercive field of approximately \(15\,\mathrm{MV\,cm^{-1}}\) and a metastable step-like feature. \textbf{B}, Evolution of non-wurtzite inversion-boundary regions along the hysteresis loop (wurtzite-like regions omitted for clarity). Narrow filaments nucleate and advance along the applied-field direction before lateral coalescence completes reversal. \textbf{C}, Populations of wurtzite-like and inversion-boundary-like environments during field cycling, showing transient accumulation of reconstructed boundary environments near switching.}}
  \label{fig:hyst}
\end{figure}

We first simulated a pristine ZnO thin film using the electric-field protocol described in Methods. The calculated polarization--field response (Fig.~\ref{fig:hyst}A) has a remanent polarization of approximately \(36\,\mu\mathrm{C\,cm^{-2}}\), a saturation polarization of \(\sim46\,\mu\mathrm{C\,cm^{-2}}\) and a coercive field of \(\sim15\,\mathrm{MV\,cm^{-1}}\). The hysteresis loop therefore reproduces the defining feature that motivates this work -- wurtzite ZnO can sustain switchable polarization, but only under very large fields. A pronounced anomaly appears near the coercive field, indicating that reversal passes through a long-lived intermediate rather than a smooth homogeneous instability. Similar loop anomalies and partially switched states have been reported experimentally in ZnMgO films \cite{yang2024coexistence,jacques2023wake}, but the atomistic sequence responsible for them has remained unclear.

{Structural snapshots identify the intermediate as a dynamically reconstructed inversion-boundary network separating oppositely polarized wurtzite regions. The boundary contains 4--8 topological motifs related to those observed in switched wurtzite nitrides \cite{calderon2023atomic,wolff2021atomic,hwang2024idb}. Time-resolved atomistic trajectories and surface-mesh representations (Supplementary Movies SM1 and SA1; Supplementary Fig.~S2) show that switching nucleates at the free surfaces (where under-coordinated atoms exist) and forms narrow, spatially rugged filaments. In the present pristine-ZnO geometry (Fig.~\ref{fig:hyst}B) the filaments initially advance rapidly along the field direction; once a sufficient number span the slab, lateral growth and coalescence complete reversal. This temporal separation accounts for the metastable step-like feature in the hysteresis loop and the transient accumulation of inversion-boundary-like environments in Fig.~\ref{fig:hyst}C.}.  This observation also hints at a two-step switching process, one lateral and one longitudinal.  As we will see, in the presence of chemical or strain disorder, the two steps are not clearly separated anymore,  reconciling why certain wurtzite systems only show this step-like feature in the measured hysteresis curves.  

{The simulated ruggedness is qualitatively consistent with the growing experimental picture that switched wurtzite boundaries need not be atomically flat or geometrically simple. In situ STEM of epitaxial (Al,B,Sc)N reports spike-like, zigzag and inclined domain boundaries, and atomic-resolution imaging reveals a structurally complex boundary region \cite{calderon2026domain}. The propagation anisotropy is not identical, the experiment finds especially rapid lateral wall motion, whereas our pristine-ZnO trajectories first develop through-thickness filaments. We therefore do not regard a particular filament geometry as universal. Instead, the common feature relevant here is a localized, reconstructed switching front with sharp features. This distinction also separates the present result from the cation-column cascades and fractal lateral fronts reported for AlN \cite{behrendt2026fractals}, our central question is which local part of a reconstructed front sets the coercive field.}

\begin{figure}[!p]
  \centering
  \includegraphics[width=\linewidth]{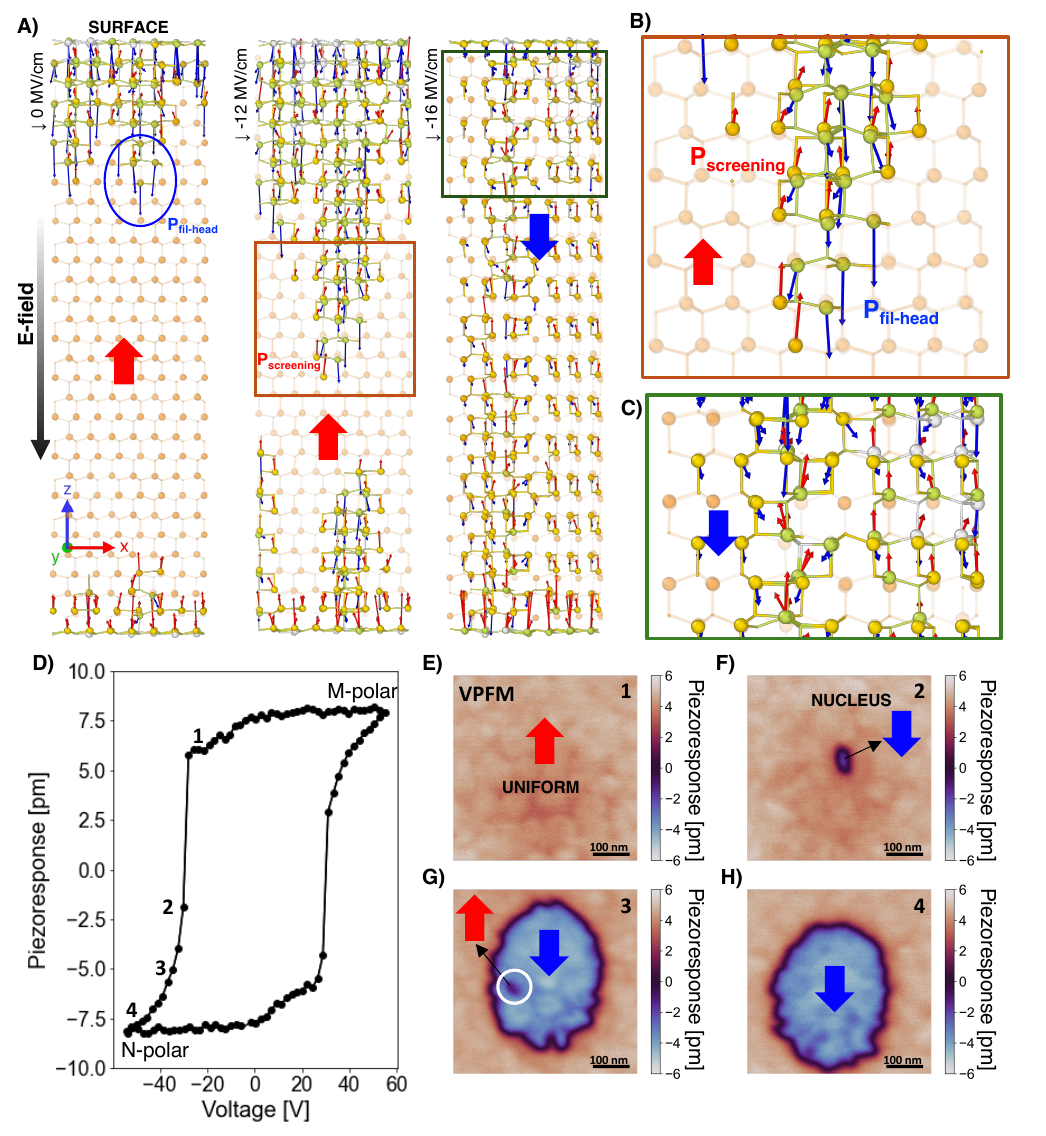}
  \caption{{\textbf{A reconstructed core--shell switching front accompanies filament growth.} \textbf{A}, Local polar-order vectors during filament growth. The advancing head contains a strongly field-aligned core response, denoted \(\mathbf{P}_{\mathrm{fil\text{-}head}}\), associated with under-coordinated cations and evolving 4--8 topological motifs; an oppositely oriented surrounding response, \(\mathbf{P}_{\mathrm{screening}}\), partially compensates the core. \textbf{B}, Enlarged view of the advancing filament head. \textbf{C}, Local structure after the filament reaches the opposite surface and the wurtzite polarity reverses. \textbf{D}, Piezoresponse-force-microscopy hysteresis response. \textbf{E--H}, PFM amplitude during field cycling showing transient coexistence and back-switching of oppositely polarized regions near reversal. The PFM observations support spatially heterogeneous switching but are not used to assign a unique atomistic filament geometry.}}
  \label{fig:filament}
\end{figure}

\subsection*{Filament-head polarization is the coercive-field descriptor}

{To determine which part of the switching front is most strongly coupled to the applied field, we analysed the site-resolved local polar-order vectors during the field cycle. At the onset of reversal, a strongly field-aligned response appears at the free surface and remains localized near the advancing inversion-boundary filament (Fig.~\ref{fig:filament}A--C). The accompanying PFM measurements (Fig.~\ref{fig:filament}D--H) show transient coexistence of oppositely polarized regions and local back-switching near reversal. We interpret these observations as evidence for spatially heterogeneous switching at few-nm length-scale with sharp atomic boundaries between up and down domains.}

{The simulations reveal that the advancing region is not a simple planar or uniformly antipolar boundary. It develops a core--shell polar-order texture: the reconstructed core at the advancing head carries a large field-aligned response, denoted \(\mathbf{P}_{\mathrm{fil\text{-}head}}\), while a surrounding tilted and oppositely oriented response, \(\mathbf{P}_{\mathrm{screening}}\), partially compensates it. In the following analysis these symbols are shorthand for features of the local charge-weighted order parameter defined in Methods and Supplementary Information.  This quantity is proportional to the local polarization in the case of uniform ZnO only, but provide a uniform metric to compare across the different perturbations to ZnO described in subsequent sections. The controlled perturbation calculations test whether changes in this head--shell local-order imbalance accompany changes in the coercive field. }


\begin{figure}[!p]
  \centering
  \includegraphics[width=\linewidth]{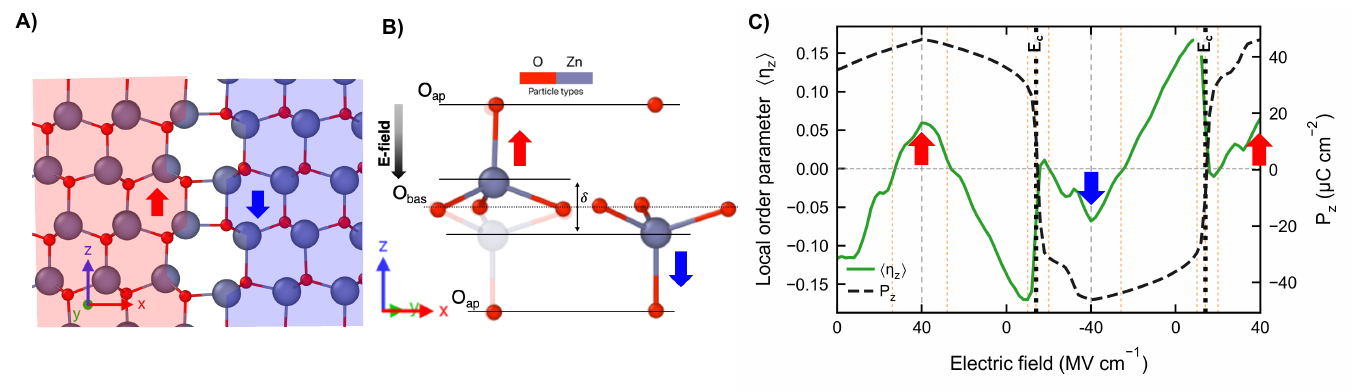}
  \caption{{\textbf{Local polar order at the advancing filament head marks the field-setting event.} \textbf{A}, Schematic inversion boundary separating oppositely polarized wurtzite regions. \textbf{B}, Local Zn displacement toward the basal oxygen plane during reconstruction; the transient geometry leaves Zn under-coordinated with respect to the apical oxygen sites. \textbf{C}, Macroscopic switching response (dashed line) and laterally averaged local polar-order parameter, \(\eta_{z,i}\) (solid line), along the field cycle. The enhanced local-order amplitude precedes the coercive-field crossing and accompanies the transient population of reconstructed inversion-boundary environments.}}
  \label{fig:undercoord-Zn}
\end{figure}

{The large head-associated response originates from the local reconstruction required to translate the inversion boundary. As sketched in Fig.~\ref{fig:undercoord-Zn}A,B, Zn moves toward and across the basal oxygen plane as the polarity reverses. Near the advancing head this motion transiently leaves metal sites under-coordinated, including threefold-coordinated Zn, while 4--8 topological motifs develop around the boundary. These configurations produce unusually large cation--anion displacement vectors, and hence large polarization at the filamentary head in pristine ZnO. The surrounding oppositely oriented shell reduces, but does not eliminate, the local head response.}

{We quantify this evolution with the site-resolved charge-weighted order parameter \(\eta_{z,i}\) defined in Methods and Supplementary Section~S1. The net and conditional averages separate changes in the population of UP/DOWN environments from changes in the magnitude of local order within each population. In pristine ZnO, the head-associated conditional amplitude rises above its saturated-wurtzite value before the macroscopic polarization crosses zero, while the population of inversion-boundary-like environments also increases near reversal (Fig.~\ref{fig:undercoord-Zn}C and Supplementary figures). The temporal association places the enhanced head response at the field-setting stage of switching rather than after complete domain reversal.}

{The conditional analysis further resolves the field-aligned head response from the oppositely oriented screening population. Switching begins as the head-associated local-order amplitude grows. Subsequent development of the opposing population increases compensation as the reconstructed filament matures. We therefore use the relative head--shell local-order response as the microscopic descriptor tested in the remainder of the paper. Importantly, we claim that models that suppress the head-associated response amplitude under the same field protocol switch at lower coercive fields. This avoids assuming that the absolute value of \(\eta_{z,i}\) is itself a thermodynamic polarization or activation energy, except in the simplest case of pristine ZnO where it is proportional to the local polarization.}

{This interpretation changes the design question from global lattice softening to local switching-front engineering. A dopant, defect or interface can lower the coercive field by reducing the local polar-order amplitude required to advance the reconstructed head, increasing the compensating response of its surroundings, or creating a site at which the same reconstructed front can nucleate at lower field. The next sections separate structural and effective-charge perturbations to test these alternatives.}

\subsection*{Charge redistribution suppresses the field-setting descriptor}

{The head--shell decomposition provides a microscopic design variable, but a useful perturbation should distinguish between structural softening and electrostatic modification. Mg substitution can change local tetrahedral geometry through size mismatch and bonding, while simultaneously modifying the effective charge distribution around the switching front. Those effects need not act in the same way. Structural disorder can broaden the distribution of local switching environments and alter boundary roughness, whereas charge redistribution can directly change the dipole moment, and hence the charge-weighted local-order response, especially at under-coordinated sites.}

{We therefore compare three controlled perturbations as three separate model systems at the same nominal modified-site fraction: strain-only, charge-only, and combined charge-plus-strain. The strain-only calculation asks how Mg-induced structural/bonding perturbations affect the switching front when the reference charge treatment is retained. The charge-only calculation keeps the Zn structural identity but changes assigned effective charges, isolating sensitivity to local electrostatics within the fixed-charge model. The combined calculation includes both contributions. These are mechanism-separation experiments, not a claim that strain and charge are fully independent degrees of freedom in a real alloy. Their purpose is to determine which perturbation most strongly changes the field-setting local-order descriptor under otherwise identical conditions.}

\begin{figure}[!p] 
    \centering 
    \includegraphics[width=0.75\linewidth]{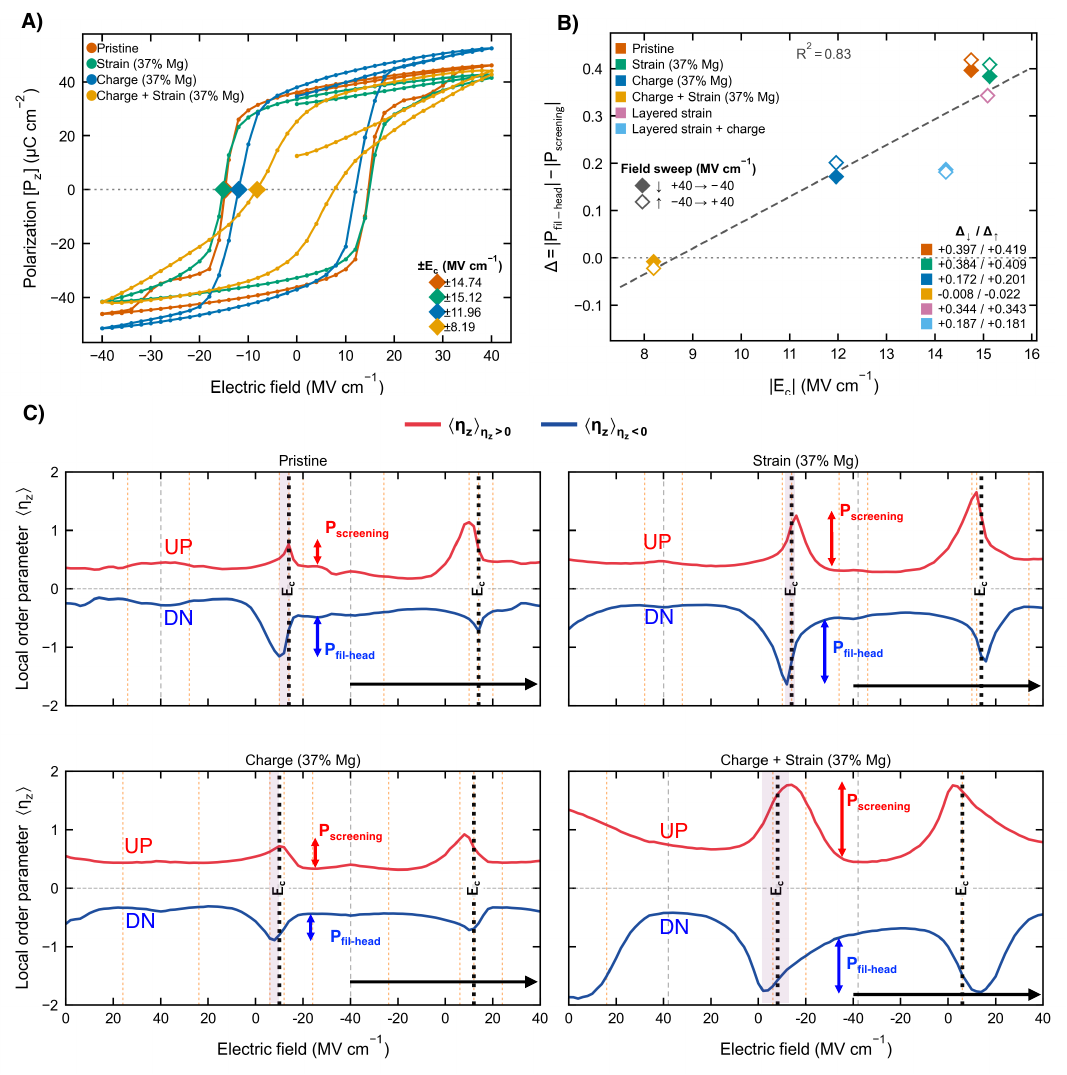} 
    \caption{\textbf{Cationic dopants and charge redistribution provide a direct route to reducing the coercive field.} \textbf{A}, Hysteresis loops for 37\% Mg substitution under three different perturbations: strain-only, charge-only, and combined strain-and-charge effects. \textbf{B}, Magnitude of the coercive field, $|E_c|$, as a function of the
local polarization imbalance, $\Delta=\bigl||P_{\mathrm{fill\text{-}head}}|-|P_{\mathrm{screening}}|\bigr|$, for the homogeneous and layered configurations. Filled and open diamonds denote the $+40\rightarrow-40$ and $-40\rightarrow+40~\mathrm{MV\,cm^{-1}}$ field sweeps, respectively. The dashed line shows a linear fit ($R^2=0.83$), revealing that smaller polarization imbalance between the filament head and its screening environment is generally associated with a lower coercive field. \textbf{C}, Conditional local order parameters for the up- and down-polarized regions, plotted along the hysteresis loop. The larger peak corresponds to the filament-head polarization, \(\mathbf{P}_{\mathrm{fil\text{-}head}}\), aligned with the applied field, whereas the smaller peak corresponds to the screening polarization, \(\mathbf{P}_{\mathrm{screening}}\), which opposes the field direction. The strain-only perturbation increases both filament-head and screening contributions but has only a limited effect on the coercive field. In contrast, when cationic substitution is accompanied by charge redistribution, the polarization at the filamentary-core head is reduced relative to the screening polarization of the surrounding shell, leading to a pronounced reduction in the coercive field. The combined strain-and-charge perturbation further lowers the coercive field, indicating that disordered strain provides an additional contribution to switching facilitation. Black horizontal arrows indicate the symmetry-related reverse-field branch of the hysteresis loop; the same features appear during field reversal in the opposite sequence, with comparable peak magnitudes.} \label{fig:mg_hyst} 
\end{figure}

In the strain-only model, 37\% Mg substitution slightly reduces remanent and saturation polarization, but it does not strongly lower the coercive field (Fig.~\ref{fig:mg_hyst}A). Instead, Mg-induced structural disorder roughens the inversion boundary, generates additional local switching sites and eliminates the pronounced metastable feature seen in pristine ZnO due to asynchronous switching of the filaments (\added{Supplementary Material movies SM2 and SA2}). The two-step switching pathway persists locally, but the nucleation and lateral expansion events are no longer synchronized across the film. This distinction is important -- local strain can reduce barriers when it creates favorable distorted environments or interfacial gradients, but random ionic-size mismatch can also roughen the switching front, weaken screening, broaden the barrier distribution or introduce disorder that resists collective propagation of switching filaments. This behavior is consistent with experiments on \(\mathrm{Zn}_{1-x}\mathrm{Mg}_x\mathrm{O}\), where robust hysteresis, wake-up, coexisting switching signatures and non-monotonic coercive-field trends suggest a competition between structural softening, charge redistribution and disorder \cite{ferri2021ferroelectrics,yang2024coexistence,spurling2025composition,baksa2024strain}. Strain from Mg therefore changes the morphology and kinetics of switching, but it does not efficiently remove the local field-setting barrier.  Local-polarization analysis as shown in Fig.~\ref{fig:mg_hyst}B shows that indeed as expected from our theoretical model above,  both the local \(\mathbf{P}_{\mathrm{fil\text{-}head}}\) in the filamentary-core and \(\mathbf{P}_{\mathrm{screening}}\) in the filamentary-shell increase by similar degrees, corresponding to the nominally smaller radii of Mg$^{2+}$ ion compared to Zn$^{2+}$ ion, leading to a negligible shift in coercive fields.

\begin{figure}[!p] 
    \centering 
    \includegraphics[width=0.75\linewidth]{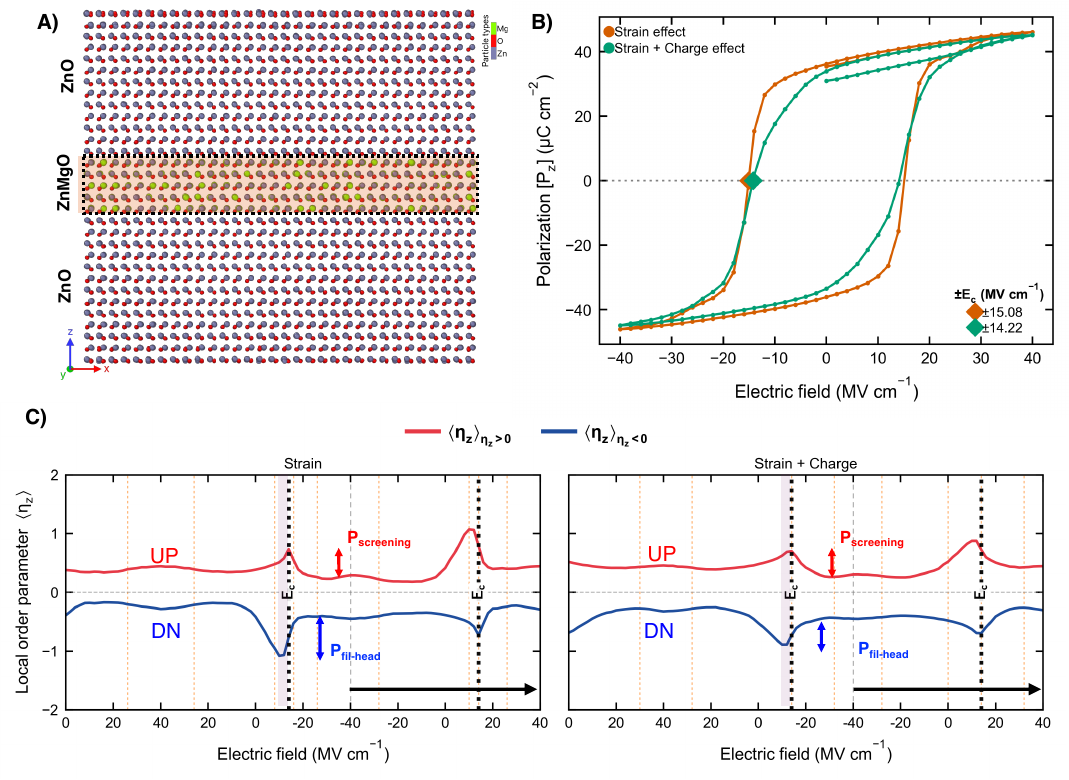} 
    \caption{\textbf{Layered dopants translate the microscopic switching mechanism into a low-field design strategy.} \textbf{A}, Schematic of the ZnO/ZnMgO/ZnO heterostructure containing a layered Mg-doped region. \textbf{B}, Hysteresis loops comparing the strain-only heterostructure with the combined charge-plus-strain heterostructure. The combined perturbation lowers the coercive field and reduces the distance over which each switching filament must propagate. \textbf{C}, Conditional average local order parameter for the strain-only and combined charge-plus-strain cases. Charge redistribution promotes filament nucleation at the buried ZnO/ZnMgO interface, enabling switching to initiate locally before propagating through the ZnO layer. The horizontal black arrow indicates the symmetry-related reverse-field branch of the hysteresis loop; the same conditional local-order features occur during field reversal in the opposite sequence, with comparable peak magnitudes.} 
    \label{fig:dopant_layered} 
\end{figure}

The charge-only model isolates the electrostatic part of the dopant response. In this model, 37\% of the metal sites were assigned an effective charge of \(+1.37e\), the remaining Zn sites carried \(+1e\), similar to pristine-ZnO and oxygen charges were adjusted to preserve neutrality uniformly by having a slightly larger negative charge of of \(-1.5069e\), while all metal atoms retained Zn masses and bonding environments. This modification lowers the coercive field to \(\sim12\,\mathrm{MV\,cm^{-1}}\), approximately \(3\,\mathrm{MV\,cm^{-1}}\) below pristine ZnO, while retaining a relatively square hysteresis loop because structural disorder is absent (Fig.~\ref{fig:mg_hyst}) (\added{Supplementary Material movies SM3 and SA3}).

{The conditional local-order analysis in Fig.~\ref{fig:mg_hyst}C explains why the charge-only perturbation is more effective than strain alone. Relative to pristine ZnO, the field-aligned peak of the charge-weighted local polar-order parameter associated with the advancing filament head decreases by nearly 50\%, whereas the oppositely oriented screening contribution changes less strongly.  This inspite of the anion charge increasing in magnitude. The peak also occurs at a lower applied field. Within the controlled fixed-charge models, the reduction in coercive field therefore co-varies with suppression of the unscreened filament-head local polar-order amplitude. (i.e. local \(\mathbf{P}_{\mathrm{fil\text{-}head}}\) - \(\mathbf{P}_{\mathrm{screening}}\) ) We interpret this trend as evidence that charge redistribution modifies the electrostatic cost of creating the advancing reconstructed head. Because the charge-only model is an effective perturbation rather than an explicit electronic-structure calculation, it is used to isolate this electrostatic contribution and not to assign a unique microscopic charge state to Mg-doped ZnO.}

{When Mg substitution and charge redistribution are combined, the coercive field decreases to approximately \(8\,\mathrm{MV\,cm^{-1}}\), nearly half the pristine-ZnO value (Fig.~\ref{fig:mg_hyst}; Supplementary Movies SM4 and SA4). The hysteresis loop is more sloped, consistent with spatially distributed switching in the structurally disordered model. At the same time, the head-associated conditional local-order amplitude is comparable to the opposing screening response, so that the unscreened filament-head local polar-order amplitude is the lowest amongst all the case. s studied here. The combination of a smaller field-setting head response and a broader distribution of nucleation environments therefore produces the largest coercive-field reduction among the uniform models.}

\subsection*{Layered ZnO/ZnMgO architectures turn the descriptor into a design rule}

The filament-head mechanism also suggests a geometric design strategy. In a homogeneous film, a filament nucleated at a free surface must travel across the full film thickness before lateral switching can complete. If a low-barrier region is placed inside the film, then filaments can nucleate from buried interfaces formed with this region forming more switching nucleation sites as well as travel shorter distances to switch the whole film. Recent experiments demonstrate that ZnO/ZnMgO heterostructures can be grown with precise control over buried interfaces and can use proximity effects to lower switching barriers in otherwise hard-to-switch polar layers \cite{skidmore2025proximity,eliseev2025proximity}. We therefore simulated a layered architecture in which a 3--6-atomic-layer Mg-rich region was embedded within a ZnO slab, creating a ZnO/ZnMgO/ZnO-like heterostructure.

{In the strain-only heterostructure, the Mg-rich layer perturbs the lattice but does not substantially reduce the head-associated local-order amplitude, and switching remains dominated by filaments nucleated near the external surfaces (Supplementary Movies SM5 and SA5). When effective charge redistribution is included, the buried ZnO/ZnMgO interface becomes an additional nucleation region (Fig.~\ref{fig:dopant_layered}B,C; Supplementary Movies SM6 and SA6), the coercive field decreases, and partially developed filaments appear inside the film. The corresponding conditional unscreened local-order response at the filament head is also reduced relative to the strain-only layered model. Because internally nucleated fronts need not traverse the full slab before contributing to coalescence, the layered architecture couples electrostatic lowering of the field-setting descriptor with geometric shortening of the propagation length. This may benefit switching kinetics as well as field, although quantitative switching times cannot be inferred from the present MD field-ramp protocol.}

\section*{Discussion}

{The simulations identify a local origin for the large coercive fields of ZnO-based wurtzite ferroelectrics and convert an atomistic switching trajectory into a materials-design principle. Recent nitride studies establish that wurtzite reversal can involve highly localized cation cascades, inversion boundaries and non-classical domain morphologies \cite{behrendt2026fractals,calderon2023atomic,lee2024switching,calderon2026domain}. Those results answer an essential pathway question. The present calculations address the next question required for materials design: once a reconstructed switching front exists, which part of it carries the field-dependent cost that controls reversal?}

{In ZnO, the largest field-aligned local polar-order amplitude develops at the advancing head of the inversion-boundary filament. The head contains under-coordinated cations and evolving 4--8 topological motifs and is surrounded by an oppositely oriented shell that partially compensates the core response. We use \(\mathbf{P}_{\mathrm{fil\text{-}head}}\) as shorthand for this field-aligned local polar-order peak; as defined in Methods and Supplementary Information, it is a charge-weighted local structural descriptor.  Across pristine, strain-only, charge-only and combined perturbations, lower coercive fields accompany suppression of this head-associated local order relative to its screening environment. This controlled comparison supports the unscreened local-order response at the filament head as the field-setting descriptor within the present model. Given that this descriptor is simply proportional to the relative distance of the anions from its metal center, it should be directly measurable in high resolution electron micrographs. Further presence of jagged switching propagation fronts seen in recent experiments on nitride-based Wurtzite ferroelectrics indicate presence of unscreened polarization at the head of these fronts.}

{This picture also rationalizes why chemically different perturbations need not have equivalent effects. Mg-induced strain and disorder roughen switching fronts and distribute nucleation events, but strain alone only weakly changes the coercive field. Effective charge redistribution acts more directly on the head--shell electrostatics and produces a larger reduction, while combining charge redistribution with Mg-induced structural disorder gives the lowest simulated coercive field, from approximately \(15\) to \(8\,\mathrm{MV\,cm^{-1}}\). Layered ZnO/ZnMgO/ZnO architectures extend the same concept geometrically by introducing buried nucleation regions and shortening the distance over which an advancing reconstructed head must propagate. The calculations therefore suggest two separable design variables: the energetic cost of the local switching front and the spatial distribution of sites from which that front can nucleate.}

{The comparison with recent microscopy is instructive but should not be interpreted as a one-to-one validation of a universal switching geometry. The atomic-scale and mesoscopic boundaries observed in epitaxial (Al,B,Sc)N are rugged, inclined and structurally complex \cite{calderon2026domain}, qualitatively resembling the finite reconstructed boundary regions that emerge in our ZnO trajectories more closely than an ideal planar-wall picture. At the same time, those experiments report faster lateral than vertical propagation, whereas our pristine free-standing ZnO slab initially develops rapid field-direction filaments., although this two-step switching blends into one when charge or dopant-strain disorder effects are incorporated. Likewise, the Rappe-group AlN calculations emphasize a sharply localized cation-column cascade and fractal lateral front \cite{behrendt2026fractals}. These differences are expected because alloy chemistry, interfaces, free surfaces, screening and field protocol all modify switching morphology. The transferable prediction of the present work is therefore not a unique geometry but the importance of the local electrostatic cost of the advancing reconstructed front.}

{The resulting design principle is actionable. Low-coercive-field ZnO-based wurtzites should be sought by (i) dopants or local chemistries that reduce the head-associated polar-order amplitude or increase its electrostatic compensation (screening), and (ii) heterostructure architectures that introduce low-barrier internal nucleation sites.   Candidate materials should therefore be evaluated not only through average lattice constants, spontaneous polarization and coherent switching barriers, but also through descriptors of the reconstructed switching front. A direct next test is to compare projected atomistic boundary structures with atomic-resolution STEM when that becomes available.  Electrode designs that mimic effects of local charging, such as spiky electrode surfaces, might also facilitate nucleation of filaments at lower coercive fields.}

{Finally, the absolute coercive fields obtained here should be interpreted in the context of the simulation protocol. Molecular-dynamics switching occurs over nanosecond-accessible timescales, at 100 K, under a discrete field ramp, and the production trajectories use fixed effective charges. Experimental coercive fields depend on temperature, frequency, thickness, electrodes, carrier screening and defect populations. The strongest conclusions of the present study therefore derive from controlled relative comparisons performed with the same cell geometry and field protocol, rather than from direct quantitative mapping of the simulated coercive fields onto a particular device measurement. Testing the proposed descriptor over temperature, field-ramp rate, dopant configuration and electronically responsive charge models will be important for establishing its quantitative transferability.}

\section*{Methods}
\begingroup

\subsection*{Reactive force field}
Reactive molecular-dynamics simulations used the Zn/Mg/O ReaxFF parameterization developed for electronically switchable polarization in \(\mathrm{Zn}_{1-x}\mathrm{Mg}_x\mathrm{O}\) \cite{sepehrinezhad2024reaxff}. ReaxFF represents bond formation and bond breaking through continuously varying bond orders and includes bonded, angular, torsional, over- and under-coordination, van der Waals and Coulombic energy terms \cite{vanduijn2001reaxff,senftle2016reaxff}. The Zn/Mg/O parameterization was trained against first-principles energetics relevant to ZnO- and MgO-derived phases and polarization reversal, including equations of state, intrinsic switching pathways and domain-wall migration pathways \cite{sepehrinezhad2024reaxff}. Reactive molecular dynamics has previously been used to resolve field-driven switching, defect effects and surface-chemistry effects in oxide ferroelectrics \cite{akbarian2019reaxff}. The purpose of the present simulations is to access the intermediate length scale between small-cell first-principles calculations and continuum descriptions, while retaining local bond reconstruction at evolving inversion boundaries.

\subsection*{Simulation cells and field-cycling protocol}
Simulations were performed with LAMMPS \cite{thompson2022lammps}. The pristine ZnO reference slab measured approximately \(9\times9\times8\,\mathrm{nm}^{3}\) and contained 55,552 atoms. A vacuum region was included normal to the slab (the crystallographic \(c\) axis, denoted \(z\)) to reduce interactions between periodically repeated surfaces and to represent a free-standing thin film. The integration time step was \(0.25\,\mathrm{fs}\). The structure was heated from 0 to 100 K over 50,000 steps and then equilibrated for a further 50,000 steps at 100 K using Berendsen temperature and pressure control, following the protocol used for the Zn/Mg/O reactive model.

After equilibration, atomic charges were held fixed during the production switching trajectories. A sawtooth electric field was applied along \(z\), spanning \(-40\) to \(+40\,\mathrm{MV\,cm^{-1}}\) in increments of \(2\,\mathrm{MV\,cm^{-1}}\); atomic configurations were recorded every \(1\,\mathrm{fs}\). The fixed-charge protocol was chosen to permit controlled separation of structural and effective-charge perturbations. It does not describe carrier injection, leakage, field-dependent electronic screening or time-dependent charge transfer. Accordingly, the charge-only and combined charge--strain calculations are interpreted as controlled effective-charge experiments rather than complete electronic-transport models. Absolute coercive fields are also expected to depend on the molecular-dynamics timescale, temperature and field-ramp protocol; comparisons among models are made under identical simulation conditions.

\subsection*{Macroscopic switching and coercive field}
The field-dependent cell dipole was obtained from the assigned atomic charges and positions. The component along the switching direction was used to construct the polarization-like hysteresis response, \(P_z\). Coercive fields were identified from the zero crossings of \(P_z\) on the forward and reverse branches of the field cycle, consistent with the analysis shown in the Supplementary Information. Because the production trajectories use assigned effective charges, \(P_z\) is used consistently as the macroscopic switching observable for comparisons among the simulated models.

\subsection*{Local structure and wurtzite-polarity assignment}
Local structural environments were analysed from the molecular-dynamics trajectories using the OVITO diamond-structure analysis. Hexagonal-diamond-like sites were then assigned a local wurtzite polarity from the geometry of their four nearest opposite-species neighbours, following the algorithm reported in the Supplementary Information. For site \(i\), bond vectors were defined as
\begin{equation}
\mathbf{b}_{ij}=\mathbf{r}_j-\mathbf{r}_i,
\end{equation}
and projected along the crystallographic \(c\) axis. The bond having the largest absolute \(c\)-axis projection was identified as the apical bond and the remaining three as basal bonds. The geometric polarity descriptor was
\begin{equation}
\Delta_i=b^{c}_{i,\mathrm{apical}}-\frac{1}{3}\sum_{j\in\mathrm{basal}}b^{c}_{ij}.
\end{equation}
For anion-centred environments the sign was reversed so that the UP/DOWN convention matched the cation-centred definition. The effective descriptor was therefore \(\Delta_i^{\mathrm{eff}}=s_i\Delta_i\), with \(s_i=+1\) for cation-centred sites and \(s_i=-1\) for anion-centred sites. Hexagonal-diamond-like environments with positive or negative \(\Delta_i^{\mathrm{eff}}\) were assigned as wurtzite-UP or wurtzite-DOWN, respectively; sites for which the local polarity could not be assigned were retained as other/non-wurtzite environments. The full assignment procedure and pseudocode are provided in the Supplementary Information.

\subsection*{Local polar-order parameter}
To quantify local polar order independently of the macroscopic hysteresis response, we use the charge-weighted site descriptor defined below, that is applicable to all types of perturbations studied in this manuscript.  We take this descriptor approach because the local dipole moment for a ZnO-based unit-cell becomes origin dependent when the unit cell is not charge-neutral, making it difficult to separate strain,  defect and electrostatic effects across the different cases.  For each cation-centred ZnO or MgO-like tetrahedron, the four nearest oxygen neighbours were identified and
\begin{equation}
\mathbf{p}_i=\sum_{j=1}^{4}q_{\mathrm{O},j}\left(\mathbf{r}_{\mathrm{O},j}-\mathbf{r}_{M,i}\right),
\end{equation}
where \(M=\mathrm{Zn}\) or Mg, \(\mathbf{r}_{M,i}\) is the central-cation position, \(\mathbf{r}_{\mathrm{O},j}\) is the position of oxygen neighbour \(j\), and \(q_{\mathrm{O},j}\) is its assigned charge. The scalar local polar-order parameter is the projection along the switching axis,
\begin{equation}
\eta_{z,i}\equiv p_{z,i}=\mathbf{p}_i\cdot\hat{\mathbf{z}}.
\end{equation}
This quantity is a site-resolved, charge-weighted structural descriptor and is not the macroscopic polarization. The net local polar order is
\begin{equation}
\langle\eta_z\rangle=\frac{1}{N}\sum_{i=1}^{N}\eta_{z,i},
\end{equation}
where \(N\) is the number of sites. Conditional averages were used to resolve the two oppositely oriented local populations,
\begin{equation}
\langle\eta_z\rangle_{\eta_z>0}=\frac{1}{N_+}\sum_{\eta_{z,i}>0}\eta_{z,i},\qquad
\langle\eta_z\rangle_{\eta_z<0}=\frac{1}{N_-}\sum_{\eta_{z,i}<0}\eta_{z,i}.
\end{equation}
These conditional amplitudes are the quantities used to identify the field-aligned filament-head response and the oppositely oriented screening response in Figs.~\ref{fig:mg_hyst} and \ref{fig:dopant_layered}. The signs of selected conditional curves are reversed only for visualization, as described in the Supplementary Information; this plotting convention does not change the underlying definitions. We use \(\mathbf{P}_{\mathrm{fil\text{-}head}}\) and \(\mathbf{P}_{\mathrm{screening}}\) in the main text as compact labels for these head- and shell-associated local polar-order responses, not as independent definitions of macroscopic polarization.  For the case of pure ZnO the local site descriptor is simply the scaled local polarization of the corresponding ZnO unit.  

\subsection*{Dopant and heterostructure perturbations}
Uniform Mg-substituted models were generated by replacing Zn with Mg at the target concentration. Three controlled perturbations were used to separate structural and electrostatic contributions. In the strain-only model, Mg substitution introduced the Mg atomic identity and associated structural/bonding perturbation while retaining the nominal reference-charge treatment. In the charge-only model, selected metal sites retained the Zn structural identity but were assigned modified effective charges. For the 37\% charge-only case, 37\% of metal sites were assigned \(+1.37e\), the remaining Zn sites \(+1e\), and oxygen charges were adjusted to maintain overall neutrality. The combined model included both Mg substitution and the effective-charge redistribution. The value \(+1.37e\) is therefore used as a controlled perturbation within the fixed-charge framework; conclusions from this model concern the sensitivity of switching to charge redistribution rather than a claim of a unique static Mg charge state.

Layered structures were constructed by introducing a 3--6-atomic-layer Mg-rich region within the ZnO slab, producing a ZnO/ZnMgO/ZnO-like architecture. Strain-only and combined charge--strain variants were subjected to the same equilibration and electric-field protocol as the uniform films. Filament nucleation, propagation, phase populations and local polar order were analysed separately in the ZnO regions and near the buried Mg-rich interfaces. Supplementary Movies SM1--SM6 and SA1--SA6 provide the corresponding surface-mesh and atomistic trajectory visualizations, and the associated per-frame field and order-parameter data are provided in the Supplementary Information.

\endgroup

\section*{Data availability}

{The data supporting the findings of this study, including molecular-dynamics trajectories and derived analysis files, will be made publicly available at [repository/DOI to be inserted before submission].}

\section*{Code availability}

{LAMMPS input files and analysis scripts used to generate the reported results will be made publicly available at [repository/DOI to be inserted before submission].}

\section*{Acknowledgements}

Molecular dynamics simulations were supported by the Center for Nanophase Materials Sciences (CNMS), a US Department of Energy, Office of Science User Facility at Oak Ridge National Laboratory. Development of the ReaxFF model and piezoresponse force microscopy was supported by the Center for 3D Ferroelectric Microelectronics (3DFeM), an Energy Frontier Research Center funded by the US Department of Energy, Office of Science, Office of Basic Energy Sciences under award DE-SC0021118, and was conducted as part of a user project at CNMS. This research used resources of the National Energy Research Scientific Computing Center (NERSC), a Department of Energy User Facility, under NERSC award BES-ERCAP35988, and resources of the Oak Ridge Leadership Computing Facility at Oak Ridge National Laboratory, supported by the Office of Advanced Scientific Computing Research, Office of Science, US Department of Energy, under contract DE-AC05-00OR22725.

\section*{Author contributions}

A.D. performed the molecular dynamics simulations and analysed the switching trajectories. A.S. and A.C.T.v.D. led development of the Zn/Mg/O ReaxFF force field and contributed to the simulation design. W.P. and J.-P.M. contributed experimental synthesis and characterization. K.P.K. contributed to experimental analysis and interpretation. B.S. contributed to interpretation of the computational results. P.G. conceived and supervised the study and led manuscript preparation. All authors discussed the results and contributed to writing the manuscript.

\section*{Competing interests}

The authors declare no competing interests.

\bibliographystyle{unsrtnat}
\bibliography{sample_ncomms_revised}

\end{document}